\documentclass[submission,copyright,creativecommons]{eptcs}
\providecommand{\event}{FESCA 2016} 

\usepackage{subfigure}
\usepackage{graphicx}
\usepackage{amssymb}
\usepackage{amsmath}
\usepackage{amsfonts}
\usepackage{array}
\usepackage[all]{xy}

\newenvironment{mcrl2}%
{\begin{trivlist}
\item\begin{tabular}{@{}>{\bf}p{2.3em}L@{\ }L@{\ }L@{\ }L@{\ }L@{\ }L@{\ }L@{\ }L}}%
{\end{tabular}\end{trivlist}}

\newcolumntype{L}{>{$}l<{$}}
 	\newcolumntype{C}{>{$}c<{$}}
 	\newcolumntype{R}{>{$}r<{$}}

\newcommand{\nat}{\mathbb{N}}
\newcommand{\ap}{{:}}

\newcommand{\List}{\it List}
\newcommand{\eqstate}{\textit{St}}

\newcommand{\plane}{\textit{Plane}}

\newcommand{\state}{\textit{PSt}}
\newcommand{\xray}{\textit{XRay}}

\newcommand{\din}{\textit{In}}
\newcommand{\dout}{\textit{Out}}

\newcommand{\xr}{\textit{xr}}
\newcommand{\pl}{\textit{pl}}
\newcommand{\condfun}{\textit{cond}}
\newcommand{\eval}{\textit{Eval}}
\newcommand{\condthenfun}{\textit{condthen}}

\newcommand{\bool}{{\mathbb B}}
\newcommand{\cons}{\ensuremath{\hspace{0.12em}\triangleright\hspace{0.08em}}}

\newcommand{\true}{\textit{true}}

\title{Industrial Experiences with a Formal DSL Semantics to Check the Correctness of Generated DSL Artifacts\thanks{This research was supported by the Dutch national COMMIT program under the Allegio project, and by the European ARTEMIS program under the Crystal project.}}
\author{Sarmen Keshishzadeh
\institute{Eindhoven University of Technology\\ Eindhoven, The Netherlands}
\email{s.keshishzadeh@tue.nl}
\and
Arjan J. Mooij
\institute{Embedded Systems Innovation by TNO\\ Eindhoven, The Netherlands}
\email{arjan.mooij@tno.nl}
\and
Jozef Hooman
\institute{Embedded Systems Innovation by TNO\\ Eindhoven, The Netherlands}
\institute{Radboud University Nijmegen\\ Nijmegen, The Netherlands}
\email{jozef.hooman@tno.nl}
}
\def\titlerunning{Checking the Correctness of Generated DSL Artifacts}
\def\authorrunning{S. Keshishzadeh, A.J. Mooij \& J. Hooman}
\begin{document}
\maketitle

\begin{abstract}
A domain specific language (DSL) abstracts from implementation details and is aligned with the way domain experts reason about a software component. The development of DSLs is usually centered around a grammar and transformations that generate implementation code or analysis models. The semantics of the language is often defined implicitly and in terms of a transformation to implementation code. In the presence of  multiple transformations from the DSL, the correctness of the generated artifacts with respect to the semantics of the DSL is a relevant issue. We  show that a formal semantics is essential for checking the correctness of the generated artifacts. We exploit the formal semantics in an industrial project and use formal techniques based on equivalence checking and model-based testing for validating the correctness of the generated artifacts. We report  about our experience with this approach in an industrial development project.
\end{abstract}

\section{Introduction}\label{sec:acq_introuduction}
A domain specific language (DSL) \cite{DKV00} abstracts from implementation details and is aligned with the way domain experts reason about a software component. By focusing on the essential concepts in a problem domain, DSLs facilitate the involvement of domain experts in the development of DSL specifications.

Tool support for the development of DSLs is improving constantly. Language workbenches such as Xtext enable language designers to define their languages and develop transformations that generate various artifacts from DSL models. This has boosted the popularity of DSL approaches in industry, as witnessed by reports like \cite{MHA13,VLHW13}.

The development of DSLs is usually centered around a grammar and transformations that generate implementation code or analysis models from DSL specifications. In such a setting, the main focus is on the transformation to implementation code which is very valuable in industrial practice. Generating analysis models is particularly interesting for safety-critical components and facilitates analyzing DSL models using a combination of formal techniques. For example, properties can be verified against verification models or simulation models can be used to explore the modeled behavior interactively.  

In DSL approaches, the semantics of the language is often defined implicitly and in terms of the generated implementation. Although DSLs focus on the essential concepts of their respective domains, the semantics of the language constructs are not always obvious. The lack of a formal semantics can give rise to different interpretations and cause inconsistencies between the transformations.

Various authors \cite{ABE11,BFLM05} have proposed to use a formal semantics to describe the precise meaning of the language constructs. The formalization also allows them
to have a single reference that should be followed by all the transformations. However, having a formal semantics as a reference does not guarantee the correctness of the transformations from a DSL. The developer of a transformation should have a deep understanding of the DSL and the target language and construct a transformation that does not deviate from the semantics of the DSL. In \cite{BG13} the authors indicate that such tasks are very error-prone and introducing redundant validations is an effective way to reduce the rate of mistakes.



We propose to use the formal semantics of a DSL and introduce redundancy to validate the correctness of the artifacts generated from transformations \cite{KM16}. We introduce redundant validations using the following formal techniques:
\begin{itemize}
  \item equivalence checking (by transforming models to a formalism that allows checking equivalence of behaviors);
  \item model-based testing (by testing conformance of executable models to a test model).
\end{itemize}

We report on our experiences with these formal techniques in the context of an industrial DSL. This DSL is used in a development project to write specifications for an existing implementation of an industrial software component and to develop new enhanced  specifications for future releases of the software. In this paper, we focus on DSL models that describe the existing implementation. The results obtained in the project show that the redundancy introduced by equivalence checking and model-based testing can effectively detect inconsistencies between the generated artifacts for a DSL. To hide the complexities of these techniques, we have  developed a push-button technology that allows industrial users to automatically perform these checks for the generated artifacts. 

Instead of validating the correctness of the generated artifacts, some authors \cite{EE08} propose to formally prove the correctness of transformations. For a realistic DSL, this can be very costly in terms of time and the required expertise \cite{L09}. Thus, proving transformations should only be considered for well-established languages and transformations. Moreover, for a young DSL, transformations are improved regularly, and hence proving their correctness may not be effective. The industrial context of our work also calls for a pragmatic approach where there is no time for time-consuming proofs.

\paragraph{Overview.}
We discuss the mCRL2 process algebra in Section~\ref{sec:acq_pre}. In Section~\ref{sec:acq_casestudy} we describe an industrial control component and informally introduce a DSL for describing its behavior; the semantics of the language is formalized  in Section~\ref{sec:acq_semantics} using mCRL2. In Section~\ref{sec:acq_mbt} we use
model-based testing to assess the quality of an implementation of the industrial component. The correctness of the used models is validated in Section~\ref{sec:acq_comparison}. In Section~\ref{sec:acq_transformations} we discuss about our experiences with different types of model transformations. In Section~\ref{sec:acq_related} we discuss related work. Section~\ref{sec:acq_conclusion} contains conclusions and  directions for future research. 

\section{Preliminaries}\label{sec:acq_pre}
In this section we give an overview of the micro Common Representation Language 2 (mCRL2) \cite{GM14}. mCRL2 is a process algebra that extends the Algebra of Communicating Processes (ACP) \cite{BK84} with data and time. The mCRL2 language and its supporting toolset \cite{mCRL2} can be used to specify and analyze the behavior of distributed systems and protocols.

In this short overview, we focus on the language constructs that we need throughout the paper. The interested reader can refer to \cite{GM14} for more details. We explain the way data types are defined and used in mCRL2 (Section~\ref{subsec:dt}) and describe behavioral specifications in the language (Section~\ref{subsec:proc_spec}).

\subsection{Data Specification}\label{subsec:dt}
mCRL2 offers ways to specify data types (also known as sorts) and use their elements in specifications. Standard data types such as natural numbers ($\nat$) and booleans ($\bool$) are predefined in the language. Common operations on these  data types are also available, e.g., $\approx$ denotes equality.  

The user can define new data types in a specification. A new sort can be declared by explicitly characterizing its elements in a structured data type. For instance, we can define ${\it Color}$ with elements ${\it Red}$, ${\it Green}$, and ${\it Blue}$:
\begin{mcrl2}
sort & {\it Color} ~ = ~ {\bf struct} ~~~{\it Red}~|~{\it Green}~|~{\it Blue};
\end{mcrl2}
It is also possible to declare structured types that depend on other sorts. For instance, a data type called ${\it Message}$ that contains pairs of natural numbers can be defined as follows:
\begin{mcrl2}
sort & {\it Message} ~ = ~ {\bf struct}~~~{\it Pair}({\it fst}\ap \nat,{\it snd}\ap \nat);
\end{mcrl2}
A ${\it Message}$ has the shape ${\it Pair}(n_1,n_2)$ where $n_1,n_2\in \nat$. The declaration of ${\it Message}$ provides two projection functions, ${\it fst}$ and ${\it snd}$, that extract the elements of ${\it Pair}(n_1,n_2)$:
\begin{align*}
{\it fst}({\it Pair}(n_1,n_2))=n_1\qquad\qquad{\it snd}({\it Pair}(n_1,n_2))=n_2
\end{align*}

Functions can also be declared and used in the mCRL2 language. Given two sorts $A$ and $B$, the notation $A\rightarrow B$ denotes the sort of functions from $A$ to $B$. mCRL2 includes an operator called function update for unary functions. For $f\in A\rightarrow B$ this operation is denoted by $f[a\rightarrow b]$ and represents a function that maps $a$ to $b$ and maps  all the other elements of $A$ like $f$ does. 

We can declare the sort of functions from $\nat$ to $\nat$ as follows:  
\begin{mcrl2}
sort & {\it NatFunc} ~ = ~ \nat \rightarrow \nat;
\end{mcrl2}
We consider two examples of this sort: 
\begin{itemize}
\item
${\it succ}$: given $n\in \nat$ returns $n+1$;
\item
${\it condsqr}$: given $n\in \nat$ returns $n^2$ if $n>10$; otherwise, it returns $n$.
\end{itemize}
We declare these functions using the ${\bf map}$ keyword:
\begin{mcrl2}
map  &~{\it succ}, {\it condsqr}~\ap~{\it NatFunc};
\end{mcrl2}
To define ${\it succ}$ and ${\it condsqr}$, it is necessary to specify the calculations performed in these functions. This is realized by introducing equations using the keyword ${\bf eqn}$. Variables used in the equations are declared by the keyword ${\bf var}$. 
\begin{mcrl2}     
var  & n~\ap~\nat;\\
eqn  & {\it succ}(n)=n+1;\\
     & {\it condsqr}(n)={\it if}(n>10,n^2,n);
\end{mcrl2}
The conditional operation has the shape ${\it if}(c,t,u)$. It evaluates to the term $t$ if the condition $c$ holds and it evaluates to the term $u$ if $c$ does not hold. 


Lists are a built-in data type in mCRL2. The set of lists where all elements are from a sort $A$ are represented by $\List(A)$. Elements of $\List(A)$ are built with two constructors: $[]$ the empty list, and $a\cons \ell$ which puts $a$ (of type $A$) in front of list $\ell$ (of type $\List(A)$). A list can be defined by specifying its elements and putting them between square brackets. For example, $[22,4]$ is a list of natural numbers. 

\subsection{Process Specification}\label{subsec:proc_spec}
mCRL2 allows us to specify behavior using a small set of primitives and operators. We use a simple example to describe some basic constructs of mCRL2. The example is a modulo $3$ counter that starts counting from $0$ and resets itself when it reaches $3$. 

In mCRL2, behavior is described in terms of processes. Actions are elementary processes and represent observable a\-tom\-ic events. Actions can also carry data parameters. In our example, ${\it count}$ and ${\it reset}$ can be considered as actions performed by the counter. The action ${\it count}$ carries a data parameter to indicate the current number. These actions are declared as follows:
\begin{mcrl2}
act & {\it count}~\ap~\nat;\\
    & {\it reset};
\end{mcrl2}

Actions can be combined using different operators to form processes that specify more complex behaviors. For instance, the non-deterministic choice between process $p$ and $q$ is denoted by $p+q$ and the sequential composition of $p$ and $q$ is denoted by $p.q$. Data values can also influence the course of actions. Suppose $c$ is a boolean expression. The process $c\rightarrow p\diamond q$ behaves as $p$ if $c$ holds and otherwise it behaves as $q$. The ``else'' part of the conditional operator can be omitted. If $c$ does not hold in $c\rightarrow p$, deadlock will occur. 

\begin{figure}
\centerline{
\xymatrix@C=4pc{
\ar[r] & *++[o][F]{} \ar[r]^{{\it count}(0)} & *++[o][F]{} \ar[r]^{{\it count}(1)} & *++[o][F]{} \ar[r]^{{\it count}(2)} & *++[o][F]{} \ar@/^1.3pc/[lll]^{{\it reset}}
}
}
\caption{Behavior of ${\it Counter}$}
\label{fig:acq_lts}
\end{figure}
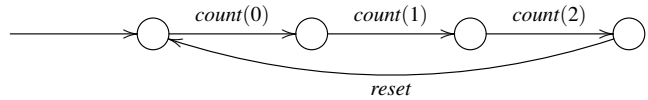

The following process specifies the modulo $3$ counter. The process ${\it Counter}$ carries a data parameter to keep track of the current number. The initial behavior is specified by ${\bf init}$, i.e., counting starts from $0$.
\begin{mcrl2}
proc & ~{\it Counter}(n\ap\nat)=(n<3)\rightarrow {\it count}(n).{\it Counter}(n+1)  \\
     &~\phantom{{\it Counter}(n\ap\nat)}+(n \approx 3)\rightarrow {\it reset}.{\it Counter}(0);\\
init & {\it Counter}(0);     
\end{mcrl2}
For any $n\leq 3$ exactly one of the conditions $n<3$ and $n\approx 3$ will evaluate to ${\it true}$. The process performs action ${\it count}(n)$ when $n<3$ and then behaves as ${\it Counter}(n+1)$. If $n\approx 3$, the process performs ${\it reset}$ and starts counting from $0$. Fig.~\ref{fig:acq_lts} depicts the labeled transition system of the counter.

\section{Industrial Application: a Clinical X-ray Generator}\label{sec:acq_casestudy}

In this section we introduce an industrial control component that we use for reporting our experiences (Section~\ref{subsec:acq_ph}). We also informally describe a DSL for specifying the behavior of the component (Section~\ref{subsec:acq_dsl}).

\subsection{Platform}\label{subsec:acq_ph}

\begin{figure}[t]
\centering
\subfigure[Interventional X-ray]
{\includegraphics[height=3.7cm]{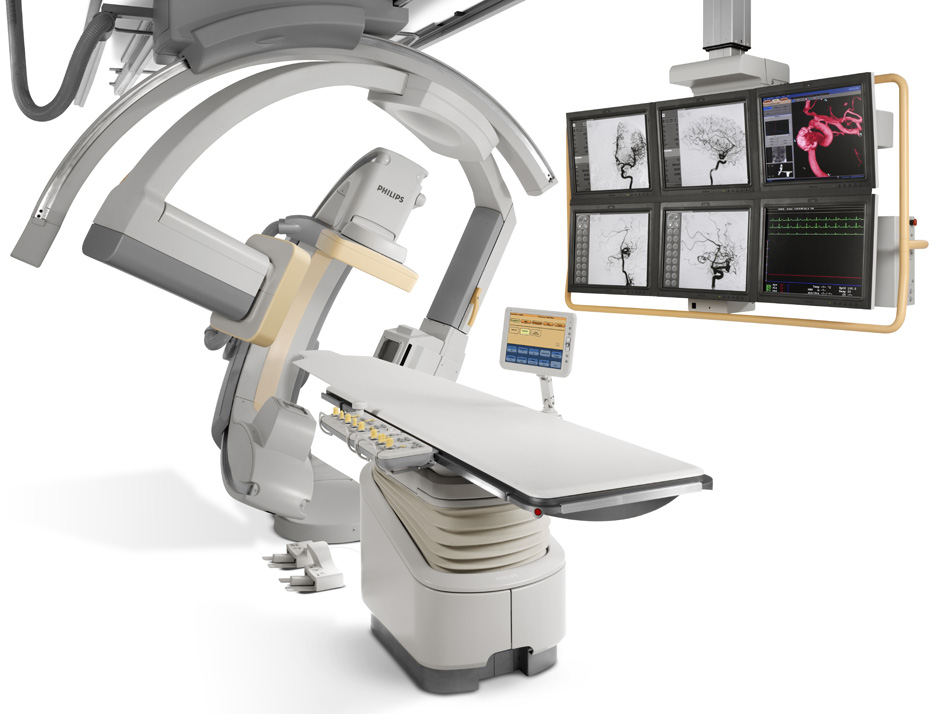}\label{fig:acq_ixr}}
\quad
\subfigure[Interfaces of \textit{Pedal Handling}]
{\includegraphics[height=3.7cm]{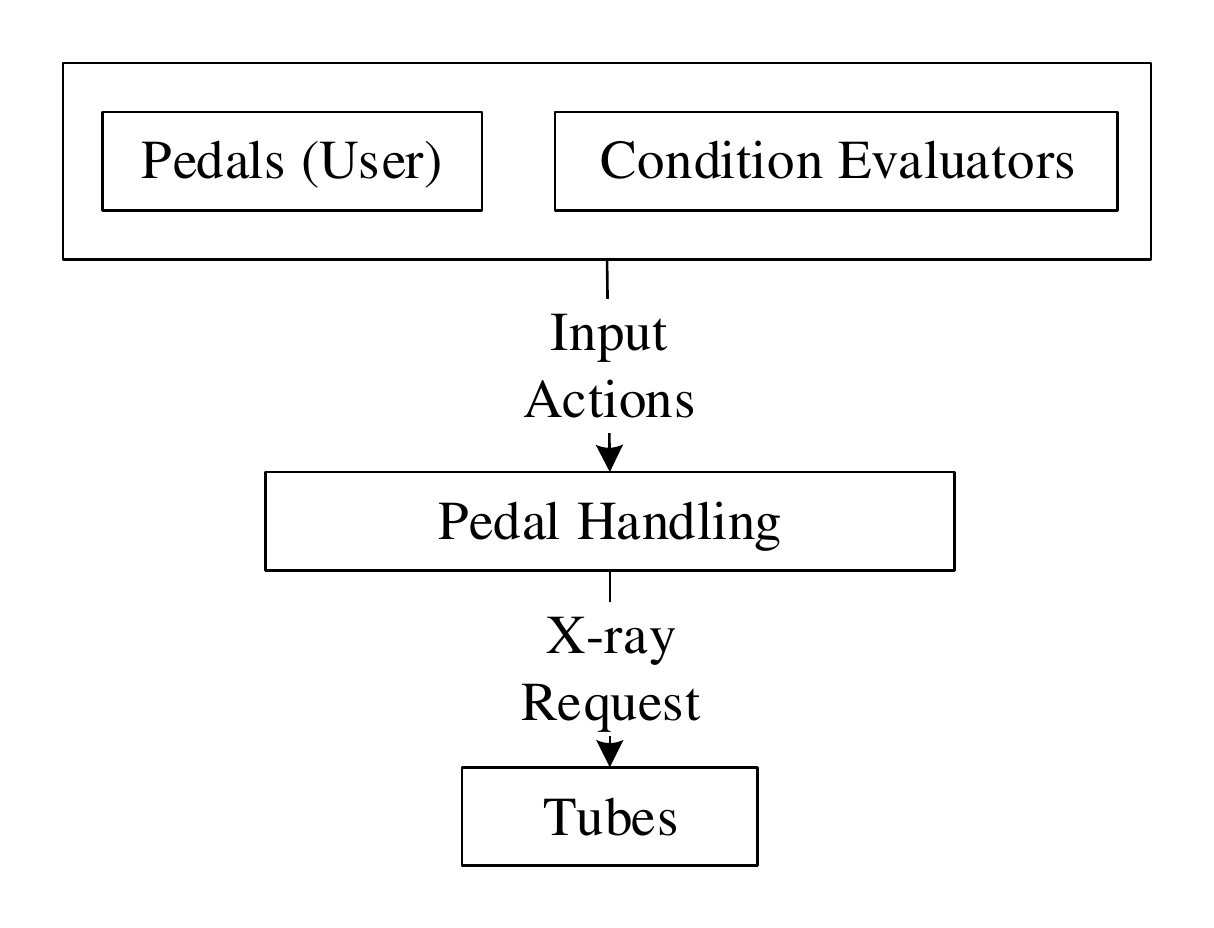}\label{fig:acq_layers}}
\caption{Industrial Application}
\end{figure}

Philips Healthcare produces interventional X-ray systems (Fig.~\ref{fig:acq_ixr}) which are used to perform minimally-invasive medical procedures. During a procedure, the surgeon uses the images on the monitors as guidance. Images are constructed for two projections. The X-ray system consists of two planes: frontal (top-down) and lateral (left-right). These planes can be used separately or together (biplane).

The surgeon sends X-ray requests using the pedals. The interventional system of Fig.~\ref{fig:acq_ixr} includes a component called \textit{Pedal Handling} that makes decisions about the amount of X-ray that should be generated in the tube of each plane (Fig.~\ref{fig:acq_layers}). From each plane the following types of X-ray can be generated:
\begin{itemize}
\item
Fluoroscopy: low dose X-ray, for obtaining real-time images;
\item
SingleShot: high dose X-ray, for capturing a single image;
\item
Series: high dose X-ray, for recording a series of images.
\end{itemize}

\textit{Pedal Handling} also takes into account conditions that should interrupt the X-ray, or that should prevent the X-ray from starting. \textit{Condition Evaluators} continuously evaluate these conditions and notify \textit{Pedal Handling} when changes occur.

\subsection{DSL for Pedal Handling}\label{subsec:acq_dsl}

To describe the behavior of \textit{Pedal Handling}, domain experts mainly focus on the external interfaces of the component. Starting from the initial state, they think about the input actions received from \textit{Pedals} and \textit{Condition Evaluators}. Based on the received input, the component might change its current state; it also makes a decision about the output X-ray and sends a request to the tubes. This process continues by receiving the next input. Thus, from the domain expert's point of view the behavior of \textit{Pedal Handling} can be described with alternating sequences of input and output actions. 

\begin{figure}[t]
\centering
\includegraphics[scale=0.55]{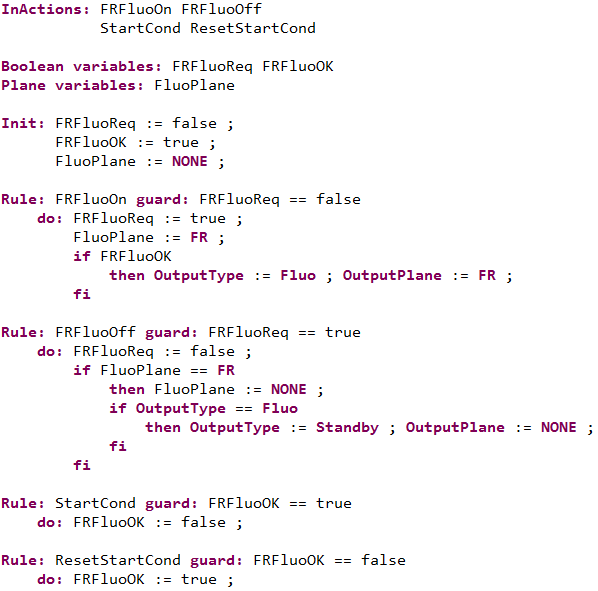}
\caption{Snapshot of a DSL Model\label{fig:acq_dslexample}}
\end{figure}

To specify the behavior of \textit{Pedal Handling}, we use a DSL that fits this way of reasoning. Fig.~\ref{fig:acq_dslexample} depicts a specification in the DSL. For confidentiality reasons, we do not provide a realistic model. This language is designed in collaboration with domain experts from Philips Healthcare. Although we use Fig.~\ref{fig:acq_dslexample} as a running example to illustrate our approach, the results reported in the paper are based on realistic DSL models and
implementations that are executable on the physical hardware. 

A DSL model starts with declaring the input actions that can be received by the \textit{Pedal Handling} component (\texttt{InActions}). Afterwards, it declares variables that keep track of the current state of the component (\texttt{Boolean variables} and \texttt{Plane variables}) and their initial values (\texttt{Init}). Two not-explicitly-declared variables, \texttt{OutputType} and \texttt{OutputPlane}, determine the output of \textit{Pedal Handling}. 

The internal logic of the component is described in terms of \texttt{Rule}s. Each rule refers to an input action and consists of a guard and a do clause. The guard describes when the input action is enabled and the do clause determines how the action influences the state of the component and the output. A DSL model specifies exactly one rule for each input action. Multiple rules for an action are not supported.


Rules of the DSL use simple constructs for describing behavior. However, the precise meaning of some constructs is not obvious. For instance, it is not obvious whether evaluating the do clause of an action can be interrupted by receiving a new input action. Since a do clause may contain multiple assignments to a variable, it is also relevant to determine when a variable assignment takes effect.

In Section~\ref{sec:acq_semantics} we give a formalization of the DSL which clearly
specifies the semantics of the language. For instance, our semantics enforces that do clauses should be interpreted in an atomic way and each assignment to a variable immediately changes its value such that the previous value is overwritten.

\section{Formalizing the DSL Semantics}\label{sec:acq_semantics}
In Section~\ref{subsec:acq_dsl} we informally introduced a DSL and motivated the need for a formal semantics. In this section, we give a formalization of the DSL semantics by introducing a transformation from the DSL to mCRL2. Similar to the \textit{Pedal Handling} DSL, a compact representation/language for describing behavior in terms of labeled transition systems is very common in many application domains. Hence, our approach for formalizing the semantics of the DSL can be applied to other DSLs. 

We describe our general transformation scheme by transforming the model of Fig.~\ref{fig:acq_dslexample} to mCRL2. Our choice of mCRL2 is motivated by the expressiveness of the language, the availability of a toolset \cite{mCRL2} that supports analysis of behavior, and
our previous experience with the language and toolset. We first discuss the required data specifications (Section~\ref{subsec:acq_domain}). Then we use process expressions to describe the behavior of
DSL models (Section~\ref{subsec:acq_semantics}). Finally, we discuss the types of analysis that we perform on DSL models (Section~\ref{subsec:acq_analysis}).

\subsection{Data Specification}\label{subsec:acq_domain}
\textbf{Plane, X-Ray Type.} As mentioned in Section~\ref{subsec:acq_ph}, X-ray can be generated from three planes (frontal, lateral, and biplane). We define the data types ${\it Plane}$ and ${\it XRay}$ to describe the planes and the type of X-ray generated from them.  
\begin{mcrl2}
sort & {\it Plane} ~ = ~{\bf struct}~~~{\it None} ~|~ {\it FR}~|~{\it LT}~|~{\it BI};\\
     & {\it XRay} ~ = ~{\bf struct}~~~{\it Standby} ~|~{\it Fluo}~|~{\it SingleShot}~|~{\it Series};
\end{mcrl2}
The combination of ${\it None}$ and ${\it Standby}$ describes a situation in which no X-ray is generated from the planes. 

\noindent\textbf{State.} A DSL model declares a set of boolean and plane variables. The valuation of these variables and the two special variables \texttt{OutputType} and \texttt{OutputPlane} determine the state of the transition system described by the DSL model. 

To describe the notion of state in mCRL2, we create two structured data types ${\it B}$ (for boolean variables) and ${\it P}$ (for plane variables). The structure of these sorts corresponds to the variable declarations of the DSL model. For the model of Fig.~\ref{fig:acq_dslexample}, $B$ and $P$ are defined as follows:
\begin{mcrl2}
sort & {\it B}~=~{\bf struct}~~{\it FRFluoReq}~|~{\it FRFluoOK};\\
     & {\it P}~=~{\bf struct}~~{\it FluoPlane};
\end{mcrl2}
To specify the valuations of boolean and plane variables, we declare the following data types:
\begin{mcrl2}
sort & {\it BVals} ~=~ {\it B}\rightarrow \bool;\\
& {\it PVals} ~=~ {\it P}\rightarrow {\it Plane};
\end{mcrl2}
Finally, the notion of state can be formalized as follows:
\begin{mcrl2}
sort & {\it PSt}~=~{\bf struct}~~{\it St}(\mathit{bs}\ap\mathit{BVals}, {\it ps}\ap{\it PVals}, {\it outType}\ap {\it XRay}, {\it outPlane}\ap {\it Plane});
\end{mcrl2}
The projection functions ${\it outType}$ and ${\it outPlane}$ extract the values of \texttt{OutputType} and \texttt{OutputPlane} from states.

In a DSL model, the initial values of the boolean and plane variables are specified by \texttt{Init}. We declare $bs_0\in {\it BVals}$ and $ps_0\in {\it PVals}$ to specify the initial values in mCRL2. We also declare $s_0\in \state$ to describe the initial state.   
\begin{mcrl2}
map & bs_0: {\it BVals};\\
    & ps_0: {\it PVals};\\
    & s_0: \state;\\
eqn & bs_0({\it FRFluoReq})={\it false};\\
    & bs_0({\it FRFluoOK})={\it true};\\
    & ps_0({\it FluoPlane})={\it None};\\
    & s_0={\it St}(bs_0,ps_0,{\it Standby},{\it None});    
\end{mcrl2}
The initial values of \texttt{OutputType} and \texttt{OutputPlane} cannot be specified by \texttt{Init} in the DSL. It is assumed that initially no X-ray is generated from the planes. Hence, we specify $s_0$ such that:
\begin{align*}
{\it outType}(s_0)={\it Standby}\qquad\qquad {\it outPlane}(s_0)={\it None}
\end{align*}

\noindent\textbf{Guard, Do Clause.} A DSL model specifies one rule for each input action. The rule of an action consists of a guard and a do clause. From Fig.~\ref{fig:acq_dslexample} one can see that a guard is a function from states to booleans. A do clause consists of a sequence of assignments/conditionals that given the current state can change the values of the variables and produce a new state. We describe guards and do clauses as follows:
\begin{mcrl2}
sort & {\it Guard} ~ = ~ {\it PSt} \rightarrow \bool;\\
     & {\it DCl} ~ = ~ \List({\it PSt} \rightarrow {\it PSt});
\end{mcrl2} 
For the DSL model of Fig.~\ref{fig:acq_dslexample} with $4$ rules, we declare the guards and do clauses $g_i,d_i$ for $1\leq i\leq 4$: 
\begin{mcrl2}
map & g_1,g_2,g_3,g_4~\ap~ {\it Guard};\\
    & d_1,d_2,d_3,d_4~\ap~ {\it DCl};
\end{mcrl2}
The calculations of each guard can be described in terms of an equation. For instance, the guard of the first rule of Fig.~\ref{fig:acq_dslexample} can be defined as follows:
\begin{mcrl2}
var & b:{\it BVals};\\
    & p:{\it PVals};\\
    & \xr:{\it XRay};\\
    & \pl:{\it Plane};\\
eqn & g_1(\eqstate(b,p,\xr,\pl))=(b({\it FRFluoReq})\approx {\it false});
\end{mcrl2}

To describe do clauses, we   specify assignments and conditionals in terms of equations. To explain this, we consider the do clause of the first rule from Fig.~\ref{fig:acq_dslexample}. This do clause contains four assignments and one conditional. 

We describe each assignment as a function that updates one of the components of a state argument $\eqstate(b,p,\xr,\pl)$. We denote the assignments of the first rule by $a_1,a_2,a_3,a_4$ based on their order of appearance:
\begin{mcrl2}
map & a_1,a_2,a_3,a_4~\ap ~ \state\rightarrow \state;\\   
eqn & a_1(\eqstate(b,p,\xr,\pl))=\eqstate(b[{\it FRFluoReq}\rightarrow \true],p,\xr,\pl);  \\
    & a_2(\eqstate(b,p,\xr,\pl))=\eqstate(b,p[{\it FluoPlane \rightarrow {\it FR}}],\xr,\pl);  \\
    & a_3(\eqstate(b,p,\xr,\pl))=\eqstate(b,p,{\it Fluo},\pl); \\
    & a_4(\eqstate(b,p,\xr,\pl))=\eqstate(b,p,\xr,{\it FR});
\end{mcrl2}
For example, the first assignment updates the values of boolean variables by setting ${\it FRFluoReq}$ to ${\it true}$. 

A conditional statement of a do clause is specified by a term of the shape ${\it if}(c,t,u)$ in equations. The conditional in the first rule of Fig.~\ref{fig:acq_dslexample} can be described as follows:
\begin{mcrl2}
map & \condfun ~\ap~\state \rightarrow \state;\\   
eqn & \condfun (\eqstate(b,p,\xr,\pl))= {\it if}(b({\it FRFluoOK}),\eval(\condthenfun,\eqstate(b,p,\xr,\pl))\\
    &\phantom{cond(\eqstate(b,p,\xr,\pl))= {\it if}(b({\it FRFluoOK}}~,\eqstate(b,p,\xr,\pl));
\end{mcrl2}
In this equation, $\eval$ is a function that evaluates a sequence of assignments or conditionals and $\condthenfun$ is the ``then'' part of the conditional (see below).

A do clause is described as a sequence of assignments/conditionals. For example, we describe the do clause of the first rule ($d_1$) as a sequence of the assignments $a_1,a_2$ and the conditional $\condfun$. The ``then'' part of a conditional is also specified as a sequence of its components. The ``then'' part of the conditional in the first rule ($\condthenfun$) is a sequence of $a_3,a_4$: 
\begin{mcrl2}
map & \condthenfun~\ap~\List(\state\rightarrow\state);\\
eqn & d_1=[a_1,a_2,\condfun];\\
    & \condthenfun=[a_3,a_4];
\end{mcrl2}

Having sequences of assignments and conditionals, it is also essential to define a function that applies a sequence of statements to a state and returns the resulting state. The following mCRL2 description defines ${\it Eval}$ for this purpose: 
\begin{mcrl2}
map & \eval~\ap~ \List (\state\rightarrow \state)\times\state \rightarrow \state;\\
var & s~\ap~\state;\\
    & f~\ap~\state\rightarrow\state;\\
    & \ell~\ap~\List(\state\rightarrow\state);\\
eqn & \eval([],s)=s;\\
    & \eval(f\cons \ell,s)=\eval(\ell,f(s));
\end{mcrl2}
The function $\eval$ is defined by specifying its effect on terms of the shape $[]$ and $f\cons \ell$ (the constructors of $\List$).

\subsection{Process Specification}\label{subsec:acq_semantics}
The \textit{Pedal Handling} component performs two types of actions: input and output. A DSL model explicitly declares a set of input actions (from \textit{Pedals} and \textit{Condition Evaluators}) by \texttt{InActions}. \textit{Pedal Handling} also performs actions $\textit{output}(\xr,p)$ to send requests for $\xr\in \xray$ to $p\in \plane$. This action is not explicitly declared in the DSL. Corresponding to the input actions and the output action we declare actions in mCRL2. For Fig.~\ref{fig:acq_dslexample} we declare:
\begin{mcrl2}
act & {\it FRFluoOn}, {\it FRFluoOff}, {\it StartCond}, {\it ResetStartCond};\\
    & {\it output}~\ap ~{\it XRay}\times {\it Plane};
\end{mcrl2}

In Section~\ref{subsec:acq_dsl}, we mentioned that domain experts  describe the behavior of \textit{Pedal Handling} with alternating sequences of input and output actions. The semantics of the DSL is aligned with this intuition. We specify the semantics of the DSL model of Fig.~\ref{fig:acq_dslexample} in terms of the following processes:
\begin{mcrl2}
proc &~ {\it P_{In}}(s\ap {\it PSt})= g_1(s) \rightarrow ({\it FRFluoOn}.{\it P_{\it Out}}({\it Eval}(d_1,s)))  \\
     &\phantom{{\it P_{In}}(s\ap {\it PSt})~}+ g_2(s) \rightarrow ({\it FRFluoOff}.{\it P_{\it Out}}({\it Eval}(d_2,s)))\\
	&\phantom{{\it P_{In}}(s\ap {\it PSt})~}+ g_3(s) \rightarrow ({\it StartCond}.{\it P_{\it Out}}({\it Eval}(d_3,s)))\\     
	&\phantom{{\it P_{In}}(s\ap {\it PSt})~}+ g_4(s) \rightarrow ({\it ResetStartCond}.{\it P_{\it Out}}({\it Eval}(d_4,s)));\\     
& {\it P_{Out}}(s\ap {\it PSt}) = {\it output}({\it outType}(s), {\it outPlane}(s)).P_{\it In}(s);\\
init & {\it P_{In}}(s_0);     
\end{mcrl2}

The process $P_{\din}$ describes the behavior of \textit{Pedal Handling} when the component is ready to receive an input. It carries a data parameter that indicates the current state. The process $P_{\din}$ uses a combination of choices and conditionals for case distinction. The guards are used as conditions in the conditional operators to determine enabled actions. Performing an input action updates the state based on the corresponding do clause. The process $P_{\dout}$ describes the behavior of \textit{Pedal Handling} when the component is ready to produce an output. In this situation, ${\it output}$ is performed and it carries the X-ray type and plane extracted from the state. Performing the output action does not influence the state.

The processes $P_{\din}$ and $P_{\dout}$ enforce alternating execution of the input and output actions. Do clauses are evaluated by $\eval$. Thus, new actions cannot be performed before a do clause is completely evaluated. Moreover, assignments have an immediate effect.

\subsection{Analyzing DSL Models}\label{subsec:acq_analysis}
For a safety critical component, it is desired to use DSL models as a single source to automatically obtain models that enable analysis using various formal techniques, e.g., verification, simulation.

To enable verification, we have automated the transformation from the DSL to mCRL2. We have used the mCRL2 toolset to generate
the state spaces of DSL models and to verify properties expressed in a variant of the modal $\mu$-calculus. A realistic DSL model declares 25 input actions and their effects
in terms of rules. The corresponding state space consists of approximately 45000 states and 350000 transitions. We have verified some safety properties against this model, e.g., ``deadlock-freedom'', ``no X-ray is generated from the planes when there is no request from the user''. The interested reader can refer to Appendix A of \cite{KMH15} for a modal $\mu$-calculus formalization of some safety properties for the DSL model of Fig.~\ref{fig:acq_dslexample}. 

The mCRL2 formalization can also be used as a reference to implement transformations to other formalisms. Having a formalized semantics helps to avoid arbitrary choices in the transformations that would give a different semantics to the constructs of the DSL. 

To enable simulation, we have implemented an automated transformation from the DSL to POOSL \cite{TFGHPV07,POOSL}. POOSL is a modeling language with a semantics expressed in terms of timed probabilistic labeled transition systems. The tools available for
POOSL allow us to simulate the modeled behavior and discuss our observations with domain experts. Due to space restrictions, we do not discuss this transformation; the interested reader can refer to \cite{KMH15} for a detailed description of the transformation. 

\begin{figure}[t]
\centering
\includegraphics[scale=0.54]{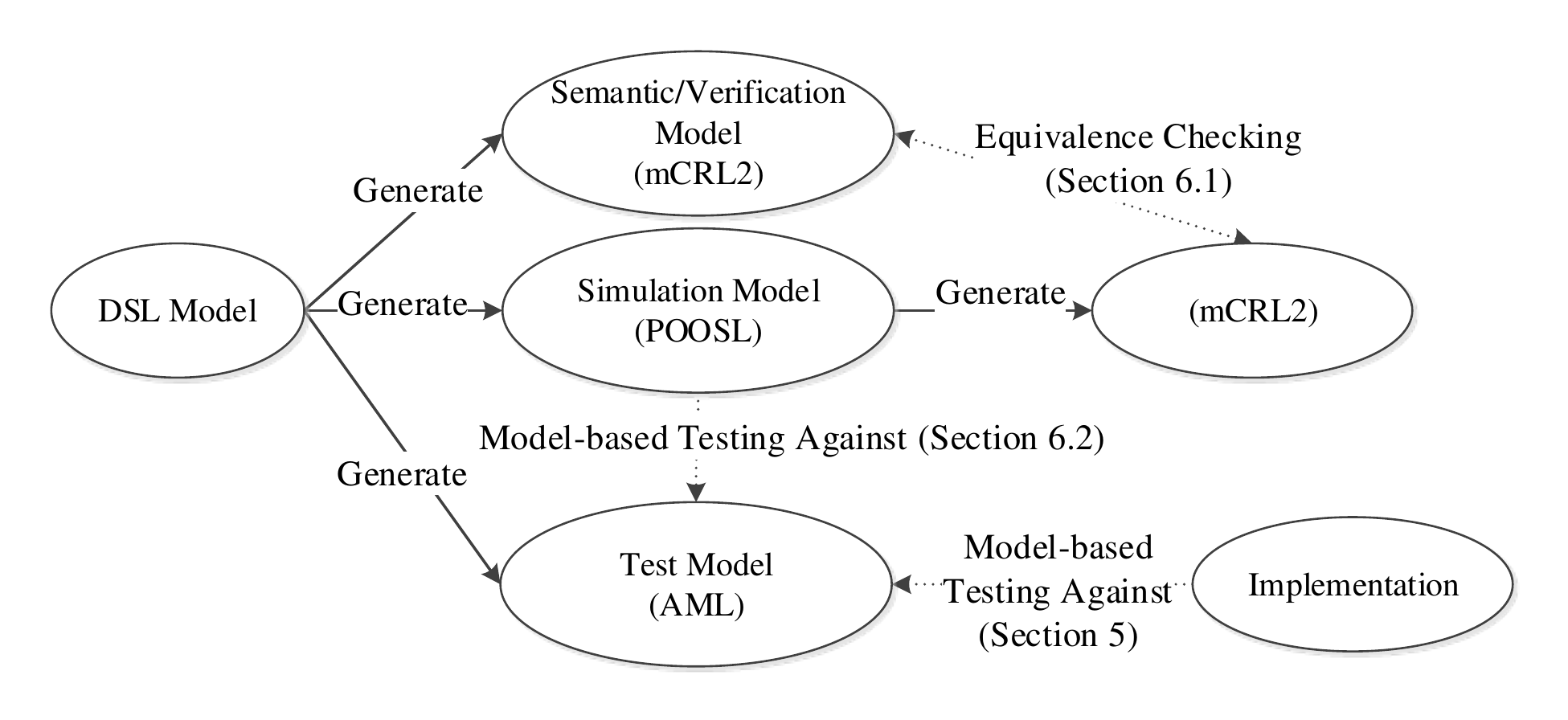}
\caption{Generated Artifacts for \textit{Pedal Handling} and their Validation\label{fig:acq_redundancy}}
\end{figure}

Using the mCRL2 formalization as guidance for implementing a transformation to
POOSL does not give a robust connection between the generated mCRL2 and POOSL models. In Section~\ref{sec:acq_comparison},
we formally validate the correctness of POOSL models with respect to the mCRL2 formalization. A proposed implementation for \textit{Pedal Handling} is available. Thus, at the moment we do not generate code from the DSL. Fig.~\ref{fig:acq_redundancy} depicts the transformations from the DSL; test models are discussed in Section~\ref{sec:acq_mbt}. In Section~\ref{sec:acq_comparison} we also discuss an approach for validating simulation models against test models.

\section{Validating the Industrial Implementation}
\label{sec:acq_mbt}
We introduce model-based testing as a way to assess the correctness of an implementation with respect to a DSL description (Section~\ref{subsec:acq_mbt}). We also discuss about interpreting
the results obtained from model-based testing (Section~\ref{subsec:acq_interpret}).

\subsection{Validating the Implementation by Model-based Testing}\label{subsec:acq_mbt}
In industry, the implementation is considered to be the most valuable artifact that is produced for a component. When behavioral models of a component are also available, it is relevant to check whether the implementation complies to the modeled behavior. Validating the compliance of implementations to DSL models adds a level of redundancy that can reveal discrepancies between the developed DSL models and implementations. 

We use model-based testing to validate the correctness of an implementation with respect to a DSL model. Model-based testing uses a model that describes the behavior of a system under test and enables both automatic generation and execution of test cases on the implementation. 

The semantics of the \textit{Pedal Handling} DSL is described in terms of labeled transition systems and hence we create a test environment based on the theory of input-output conformance (\textit{ioco}) testing for labeled transition systems \cite{T08} to automatically derive test cases from the behavior described in the DSL and execute them on the implementation. 

In the \textit{ioco} theory, correctness of implementations with respect to specifications is expressed in terms of the binary relation of \textit{ioco}. The \textit{ioco} theory provides an algorithm that derives a set of test cases from a given specification such that executing this set of test cases on an implementation determines whether the specification and implementation are related by the conformance relation. 

The model-based testing tool of Axini \cite{Axini} is based on the \textit{ioco} theory. In this tool, specifications are described in the Axini Modeling Language (AML). We have developed an
automated transformation from the DSL to AML (Fig.~\ref{fig:acq_redundancy}). The generated AML model for a DSL specification is used for model-based testing against the implementation; see \cite{KMH15} for details about the transformation to AML.

\subsection{Interpreting the Results of Model-based Testing}\label{subsec:acq_interpret}
Validating the compliance of an implementation to a DSL model by model-based testing requires special attention to interpret the results correctly. A failed test case shows a discrepancy between the test model and the implementation. This can have three different reasons:
\begin{itemize}
\item
a mistake in the transformation from the DSL to the test models;
\item
a failure in the implementation;
\item
a modeling mistake in the DSL model, which is cascaded to the test model.
\end{itemize}
The transformation from the DSL to test models should preserve the semantics of the DSL. A mistake in realizing the semantics may result in failed test cases. In Section~\ref{sec:acq_comparison} we introduce an approach to gain confidence in the correctness of the generated models in Fig.~\ref{fig:acq_redundancy}. When there is sufficient confidence in the correctness of test models, a failed test case may indicate a failure of the implementation.

The last item mentioned above is particularly relevant if implementations are not generated from the DSL. In such cases, failed test cases could also indicate a mistake in DSL models; the intended behavior is not correctly described in the DSL (and the same modeling mistakes are cascaded to the test model) but the implementation has correctly realized the behavior.

\paragraph{Results.}
For our model-based testing experiments, we used a real but not-yet-released implementation of \textit{Pedal Handling}. Independently of our experiments with the DSL, the developers have constructed an extensive suite of unit tests for validating the implementation. However, in model-based testing the focus is on assessing the observable behavior of the component based on a model by providing stimuli at its external interfaces. This    revealed a number of issues in the implementation that are out of the scope of the unit tests. For example, one of the requirements of \textit{Pedal Handling} indicates that high-dose X-ray requests have priority over low-dose X-ray requests. A design mistake in realizing this requirement led to a failed test case. The failed test was part of an unlikely trace where three different pedals must be pressed at the same time.

Model-based testing revealed that certain modeling choices taken in our DSL models are implemented differently in the implementation. Unlike stopping \emph{Fluo} (modeled by \textit{FRFluoOff} in the example of Fig.~\ref{fig:acq_dslexample}), stopping \emph{Series} requires performing two actions in a specific order. The output specified for the first step of stopping \emph{Series} was different from the output produced by the implementation. This observation led to minor changes in our DSL models.

\section{Checking the Correctness of Generated Models through Redundancy}
\label{sec:acq_comparison}
Transformations from a DSL to analysis models allow the user to apply various formal techniques and reason about DSL models. Analysis models can also be used to assess the correctness of other artifacts available for a component (Section~\ref{sec:acq_mbt}). However, the results obtained from analysis models are only valuable if the corresponding transformations correctly realize the semantics of the DSL.  

Developing a transformation from a DSL to a modeling language requires a deep understanding of the semantics of the DSL and the target language. Moreover, the transformation should not deviate from the semantics of the DSL. Introducing redundancy is a very effective way to reduce the rate of mistakes in such error-prone tasks \cite{BG13}.


We introduce redundancy to validate the correctness of the artifacts depicted in Fig.~\ref{fig:acq_redundancy} in two ways: equivalence checking (Section~\ref{subsec:acq_red_eqn}) and model-based testing (Section~\ref{subsec:acq_red_mbt}). Note that the validation techniques introduced in this section are not bound to the transformations of Fig.~\ref{fig:acq_redundancy} and hence manually constructed models can also be validated using equivalence checking and model-based testing.

%

\subsection{Checking the Behavioral Equivalence between Artifacts}\label{subsec:acq_red_eqn}
The verification and simulation models in Fig.~\ref{fig:acq_redundancy} have an underlying labeled transition system. To get confidence in the correctness of simulation models with respect to the formalized semantics (transformation to verification models), we can investigate whether the labeled transition systems described by the simulation and verification models are related by an equivalence relation, e.g., strong bisimulation. This may require to develop transformations from analysis models to a formalism that enables state space generation and comparison. 

We have used the mCRL2 toolset for state space generation (the \texttt{lps2lts} tool) and comparison (the \texttt{ltscompare} tool). To enable state space generation for simulation models, we have developed a transformation from
POOSL to mCRL2. To avoid bridging wide semantic gaps between POOSL and mCRL2, we have restricted our simulation models to a sufficient subset of POOSL (see \cite{KMH15} for more details about the constructs used in simulation models) and developed a transformation to mCRL2 for that specific subset.

When comparing behaviors, the internal steps performed by them are not relevant; we focus on the observable behaviors. Hence, we check whether the state spaces are equivalent modulo branching bisimulation. Fig.~\ref{fig:acq_redundancy} depicts equivalence checking between mCRL2 and POOSL models and the automated transformation from POOSL to mCRL2 that enables this check. 

\paragraph{Results.}
For realistic DSL models, the behaviors described in mCRL2 and POOSL were equivalent modulo branching bisimulation. Models generated from three transformations (DSL to mCRL2, DSL to POOSL, and POOSL to mCRL2) are involved in equivalence checking between simulation and verification models. Each transformation is implemented by a different person. This reduces the probability of identical mistakes in the transformations and makes the redundancy introduced by equivalence checking more valuable. 

\subsection[Model-based Testing of Executable Models]{Model-based Testing of Executable Models}
\label{subsec:acq_red_mbt}

In Section~\ref{sec:acq_mbt} we used model-based testing to validate the correctness of an implementation. Executable analysis models can also be treated as black-boxes that interact via their interfaces; we can supply inputs to an executable model and observe its output. Thus, executable models can be tested against the test model generated from a DSL model using model-based testing. Failed test cases reveal mistakes in the transformations. In our case study, we have applied model-based testing to POOSL models; see Fig.~\ref{fig:acq_redundancy}. 

Note that there are also other possibilities for validating the artifacts of Fig.~\ref{fig:acq_redundancy} (e.g., model-based testing mCRL2 models against AML models). Adding such validations  gives more confidence in the correctness of the artifacts but would require additional effort to create the required environment. 

\paragraph{Results.}
We did not encounter any failed test cases in our model-based testing experiments against POOSL models. Similar to Section~\ref{subsec:acq_red_eqn}, the transformations to simulation models (POOSL) and test models (AML) are implemented by different people. Absence of failed test cases gives more confidence that the semantics of the DSL is realized correctly in these models.

\section{Experiences with Different Kinds of Model Transformations}\label{sec:acq_transformations}
The approach of Fig.~\ref{fig:acq_redundancy} deploys model transformations to enable the use of multiple formal techniques and to introduce redundant mechanisms for assessing the correctness of different artifacts with respect to the formalized semantics. This approach relies on two types of transformation: transformations from the DSL to a general-purpose modeling language (from the DSL to mCRL2, POOSL, and AML), and transformations between general-purpose modeling languages (from POOSL to mCRL2). In this section we report on our experiences with these two types of transformations. 

General-purpose modeling languages originate from different disciplines and are applicable to a wide range of problems. For instance, POOSL is designed to be applicable for simulation and performance analysis, whereas mCRL2 is focused on formal verification. In our experience, making transformations between two general-purpose formalisms is not a trivial task. There are usually language constructs from the source language that are difficult or even impossible to translate to the target language. 
 
For example, the data layer of POOSL is object-oriented. Each data class describes its variables and methods.  Moreover, instances of certain data structures (e.g., strings, lists) are accessed using pointers. On the other hand, mCRL2 is not object-oriented and does not support pointers. Hence, transforming data classes or pointers from POOSL would require complex mechanisms in mCRL2 models. 

Similarities between the source and target languages may suggest a direct mapping between certain constructs. However, similar constructs in two formalisms may have subtly different semantics. For instance, one would expect the conditional statement \texttt{if c then p else q fi} of POOSL to be trace equivalent to $c\rightarrow p\diamond q$ in mCRL2. Based on the formal semantics of POOSL, \texttt{if c then p else q fi} first performs an internal step ($\tau$ transition) to evaluate the condition. Then it behaves as \texttt{p} if \texttt{c} holds and otherwise it behaves as \texttt{q}. However, in mCRL2 no internal action is performed for evaluating $c$ and hence the two conditional statements are not trace equivalent. This observation reiterates the importance of formal semantics in transformations
between languages.

In the literature, some authors have reported similar experiences about transformations between general-purpose modeling languages. A partial transformation from the hybrid modeling formalism Chi 2.0 \cite{BHRRS08} to mCRL2 is proposed in \cite{S12}. The intention is to enable formal verification on models in Chi 2.0. Semantic differences between Chi 2.0 and mCRL2 makes the transformation complex and hard to maintain. The generated models are also complex and sometimes difficult to analyze by tools. 

In our experience, restricting the scope of a transformation (e.g., using predefined data types in POOSL specifications instead of representing POOSL data classes in mCRL2) and studying the semantics of the relevant constructs from the source and target languages are effective ways to overcome the challenges faced in implementing transformations between general-purpose languages.  

The \textit{Pedal Handling} DSL is defined for a narrow domain and its semantics is less elaborate compared with POOSL, mCRL2, and AML. Thus, it requires less effort to construct the transformations from the DSL to different analysis models.

\section{Related Work}\label{sec:acq_related}
In \cite{ABE11} the authors prototype the semantics of a DSL called SLCO in terms of a transformation to an intermediate language called CS. Afterwards, CS models are transformed to labeled transition systems and are inspected manually or analyzed by existing tools. The authors have also implemented a number of transformations from SLCO to SLCO models with equivalent behaviors and suggested that the prototype semantics can be used to compare the underlying labeled transition systems of the source and target models.

The B method has been used in \cite{BFLM05} to develop process schedulers based on specifications in a DSL. The information given by a DSL model is taken into account at several refinement steps in B machines. The authors also introduce a decidable logic for expressing proof obligations of the refinement steps. This allows them to automatically prove the refinements.

The mentioned studies validate refinement steps, whereas we offer various formal techniques to assess the semantic correctness of different types of artifacts in an automated way.

In \cite{SBSM14} a combination of model-based techniques is used to develop a software bus in a two-phase process. In the first phase, an mCRL2 model of the component is created and validated through simulation. After developing the component, the mCRL2 model is used for model-based testing of the implementation. In the second phase, different properties are verified against the mCRL2 model. The model is improved based on the results and is used for model-based testing against a second implementation. In comparison, our approach is centered around domain-specific models and artifacts generated from them.


\section{Conclusions}\label{sec:acq_conclusion}
A DSL allows us to use models that are naturally aligned with the way domain experts reason about a software component. Existing tools enable language designers to define DSLs and to construct transformations to implementation code and analysis models. However, the semantic correctness of the generated artifacts is usually overlooked.

To resolve conflicting interpretations of a DSL, we use a formal semantics of the language. We also propose to have additional mechanisms to validate the correctness of the generated artifacts with respect to the semantics of the DSL. 


We have experimented with this approach as preparation for the redesign of a clinical X-ray generator. In this paper, we reported on our experiences with DSL models for an existing implementation of the industrial component. At the moment, the DSL and the transformation to simulation models are frequently used for discussions on potential enhancements in the behavior of the component. 

We plan to extend our approach by other forms of redundancy to increase the reliability of the artifacts. To this end, we consider using model learning. Model learning techniques extract an automaton model for an implementation by systematically performing tests on it and observing its behavior \cite{A87}. A learned model can give insight in the implemented behavior, and can be compared with the labeled transition system described by a DSL model. 

Our approach for validating the generated artifacts for a single DSL model can be extended to validate the transformations themselves based on a set that consists of several DSL models. Various criteria can also be defined for the considered sets to obtain DSL models that examine different aspects of the transformations. 

\bibliographystyle{eptcs}
\bibliography{acqcontrol}

\begin{thebibliography}{10}
\providecommand{\bibitemdeclare}[2]{}
\providecommand{\surnamestart}{}
\providecommand{\surnameend}{}
\providecommand{\urlprefix}{Available at }
\providecommand{\url}[1]{\texttt{#1}}
\providecommand{\href}[2]{\texttt{#2}}
\providecommand{\urlalt}[2]{\href{#1}{#2}}
\providecommand{\doi}[1]{doi:\urlalt{http://dx.doi.org/#1}{#1}}
\providecommand{\bibinfo}[2]{#2}

\bibitemdeclare{inproceedings}{ABE11}
\bibitem{ABE11}
\bibinfo{author}{S.~\surnamestart Andova\surnameend}, \bibinfo{author}{M.G.J.
  \surnamestart van~den Brand\surnameend} \& \bibinfo{author}{L.~\surnamestart
  Engelen\surnameend} (\bibinfo{year}{2011}): \emph{\bibinfo{title}{Prototyping
  the Semantics of a {DSL} using {ASF}+{SDF}: Link to Formal Verification of
  {DSL} Models}}.
\newblock In: {\sl \bibinfo{booktitle}{Proceedings of {AMMSE'11}}}, {\sl
  \bibinfo{series}{EPTCS}}~\bibinfo{volume}{56}, pp. \bibinfo{pages}{65--79},
  \doi{10.4204/EPTCS.56.5}.

\bibitemdeclare{article}{A87}
\bibitem{A87}
\bibinfo{author}{D.~\surnamestart Angluin\surnameend} (\bibinfo{year}{1987}):
  \emph{\bibinfo{title}{Learning regular sets from queries and
  counterexamples}}.
\newblock {\sl \bibinfo{journal}{Information and computation}}
  \bibinfo{volume}{75}(\bibinfo{number}{2}), pp. \bibinfo{pages}{87--106},
  \doi{10.1016/0890-5401(87)90052-6}.

\bibitemdeclare{misc}{Axini}
\bibitem{Axini}
\bibinfo{author}{\surnamestart Axini\surnameend}: \emph{\bibinfo{title}{{\tt
  http://www.axini.nl}}}.

\bibitemdeclare{article}{BHRRS08}
\bibitem{BHRRS08}
\bibinfo{author}{D.A. \surnamestart van Beek\surnameend}, \bibinfo{author}{A.T.
  \surnamestart Hofkamp\surnameend}, \bibinfo{author}{M.A. \surnamestart
  Reniers\surnameend}, \bibinfo{author}{J.E. \surnamestart Rooda\surnameend} \&
  \bibinfo{author}{R.R.H. \surnamestart Schiffelers\surnameend}
  (\bibinfo{year}{2008}): \emph{\bibinfo{title}{Syntax and formal semantics of
  Chi 2.0}}.
\newblock {\sl \bibinfo{journal}{Eindhoven University of Technology, Technical
  Report}}.

\bibitemdeclare{article}{BK84}
\bibitem{BK84}
\bibinfo{author}{J.A. \surnamestart Bergstra\surnameend} \&
  \bibinfo{author}{J.W. \surnamestart Klop\surnameend} (\bibinfo{year}{1984}):
  \emph{\bibinfo{title}{Process algebra for synchronous communication}}.
\newblock {\sl \bibinfo{journal}{Information and control}}
  \bibinfo{volume}{60}(\bibinfo{number}{1}), pp. \bibinfo{pages}{109--137},
  \doi{10.1016/S0019-9958(84)80025-X}.

\bibitemdeclare{inproceedings}{BFLM05}
\bibitem{BFLM05}
\bibinfo{author}{J.P. \surnamestart Bodeveix\surnameend},
  \bibinfo{author}{M.~\surnamestart Filali\surnameend},
  \bibinfo{author}{J.~\surnamestart Lawall\surnameend} \&
  \bibinfo{author}{G.~\surnamestart Muller\surnameend} (\bibinfo{year}{2005}):
  \emph{\bibinfo{title}{Formal methods meet domain specific languages}}.
\newblock In: {\sl \bibinfo{booktitle}{Proceedings of {IFM'05}}},
  \bibinfo{organization}{Springer}, pp. \bibinfo{pages}{187--206},
  \doi{10.1007/11589976\_12}.

\bibitemdeclare{article}{BG13}
\bibitem{BG13}
\bibinfo{author}{M.G.J. \surnamestart van~den Brand\surnameend} \&
  \bibinfo{author}{J.F. \surnamestart Groote\surnameend}
  (\bibinfo{year}{2013}): \emph{\bibinfo{title}{Software Engineering:
  Redundancy is Key}}.
\newblock {\sl \bibinfo{journal}{Science of Computer Programming}}
  \bibinfo{volume}{97}, pp. \bibinfo{pages}{75--81},
  \doi{10.1016/j.scico.2013.11.020}.

\bibitemdeclare{article}{DKV00}
\bibitem{DKV00}
\bibinfo{author}{A.~\surnamestart van Deursen\surnameend},
  \bibinfo{author}{P.~\surnamestart Klint\surnameend} \&
  \bibinfo{author}{J.~\surnamestart Visser\surnameend} (\bibinfo{year}{2000}):
  \emph{\bibinfo{title}{{Domain-Specific Languages}: an annotated
  bibliography}}.
\newblock {\sl \bibinfo{journal}{SIGPLAN Notices}}
  \bibinfo{volume}{35}(\bibinfo{number}{6}), pp. \bibinfo{pages}{26--36},
  \doi{10.1145/352029.352035}.

\bibitemdeclare{inproceedings}{EE08}
\bibitem{EE08}
\bibinfo{author}{H.~\surnamestart Ehrig\surnameend} \&
  \bibinfo{author}{C.~\surnamestart Ermel\surnameend} (\bibinfo{year}{2008}):
  \emph{\bibinfo{title}{Semantical Correctness and Completeness of Model
  Transformations Using Graph and Rule Transformation}}.
\newblock In: {\sl \bibinfo{booktitle}{Proceedings of ICGT'08}}, {\sl
  \bibinfo{series}{LNCS}} \bibinfo{volume}{5214},
  \bibinfo{publisher}{Springer-Verlag}, pp. \bibinfo{pages}{194--210},
  \doi{10.1007/978-3-540-87405-8\_14}.

\bibitemdeclare{phdthesis}{S12}
\bibitem{S12}
\bibinfo{author}{Stappers \surnamestart F.P.M.\surnameend}
  (\bibinfo{year}{2012}): \emph{\bibinfo{title}{Bridging Formal Models: An
  Engineering Perspective}}.
\newblock Ph.D. thesis, \bibinfo{school}{Eindhoven University of Techonology}.

\bibitemdeclare{book}{GM14}
\bibitem{GM14}
\bibinfo{author}{J.F. \surnamestart Groote\surnameend} \& \bibinfo{author}{M.R.
  \surnamestart Mousavi\surnameend} (\bibinfo{year}{2014}):
  \emph{\bibinfo{title}{Modeling and Analysis of Communicating Systems}}.
\newblock \bibinfo{publisher}{MIT press}.

\bibitemdeclare{article}{KM16}
\bibitem{KM16}
\bibinfo{author}{S.~\surnamestart Keshishzadeh\surnameend} \&
  \bibinfo{author}{A.J. \surnamestart Mooij\surnameend} (\bibinfo{year}{2016}):
  \emph{\bibinfo{title}{Formalizing and Testing the Consistency of DSL
  Transformations}}.
\newblock {\sl \bibinfo{journal}{Formal Aspects of Computing (in press)}},
  \doi{10.1007/s00165-016-0359-1}.

\bibitemdeclare{article}{KMH15}
\bibitem{KMH15}
\bibinfo{author}{S.~\surnamestart Keshishzadeh\surnameend},
  \bibinfo{author}{A.J. \surnamestart Mooij\surnameend} \&
  \bibinfo{author}{J.~\surnamestart Hooman\surnameend} (\bibinfo{year}{2015}):
  \emph{\bibinfo{title}{Industrial Experiences with a Formal {DSL} Semantics to
  Check Correctness of {DSL} Transformations}}.
\newblock {\sl \bibinfo{journal}{\href{http://arxiv.org/abs/1511.08049}{arXiv
  preprint:1511.08049}}}.

\bibitemdeclare{article}{L09}
\bibitem{L09}
\bibinfo{author}{X.~\surnamestart Leroy\surnameend} (\bibinfo{year}{2009}):
  \emph{\bibinfo{title}{Formal verification of a realistic compiler}}.
\newblock {\sl \bibinfo{journal}{Communications of the ACM}}
  \bibinfo{volume}{52}(\bibinfo{number}{7}), pp. \bibinfo{pages}{107--115},
  \doi{10.1145/1538788.1538814}.

\bibitemdeclare{misc}{mCRL2}
\bibitem{mCRL2}
\bibinfo{author}{\surnamestart mCRL2\surnameend}: \emph{\bibinfo{title}{{\tt
  http://mcrl2.org}}}.

\bibitemdeclare{inproceedings}{MHA13}
\bibitem{MHA13}
\bibinfo{author}{A.J. \surnamestart Mooij\surnameend},
  \bibinfo{author}{J.~\surnamestart Hooman\surnameend} \&
  \bibinfo{author}{R.~\surnamestart Albers\surnameend} (\bibinfo{year}{2013}):
  \emph{\bibinfo{title}{Gaining Industrial Confidence for the Introduction of
  {Domain-Specific Languages}}}.
\newblock In: {\sl \bibinfo{booktitle}{Proceedings of IEESD'13}},
  \bibinfo{publisher}{IEEE}, pp. \bibinfo{pages}{662--667},
  \doi{10.1109/COMPSACW.2013.83}.

\bibitemdeclare{misc}{POOSL}
\bibitem{POOSL}
\bibinfo{author}{\surnamestart POOSL\surnameend}: \emph{\bibinfo{title}{{\tt
  http://poosl.esi.nl}}}.

\bibitemdeclare{article}{SBSM14}
\bibitem{SBSM14}
\bibinfo{author}{M.~\surnamestart Sijtema\surnameend},
  \bibinfo{author}{A.~\surnamestart Belinfante\surnameend},
  \bibinfo{author}{M.I.A. \surnamestart Stoelinga\surnameend} \&
  \bibinfo{author}{L.~\surnamestart Marinelli\surnameend}
  (\bibinfo{year}{2014}): \emph{\bibinfo{title}{Experiences with formal
  engineering: Model-based specification, implementation and testing of a
  software bus at {Neopost}}}.
\newblock {\sl \bibinfo{journal}{Science of computer programming}}
  \bibinfo{volume}{80}, pp. \bibinfo{pages}{188--209},
  \doi{10.1016/j.scico.2013.04.009}.

\bibitemdeclare{inproceedings}{TFGHPV07}
\bibitem{TFGHPV07}
\bibinfo{author}{B.D. \surnamestart Theelen\surnameend},
  \bibinfo{author}{O.~\surnamestart Florescu\surnameend},
  \bibinfo{author}{M.C.W. \surnamestart Geilen\surnameend},
  \bibinfo{author}{J.~\surnamestart Huang\surnameend}, \bibinfo{author}{P.H.A.
  \surnamestart van~der Putten\surnameend} \& \bibinfo{author}{J.P.M.
  \surnamestart Voeten\surnameend} (\bibinfo{year}{2007}):
  \emph{\bibinfo{title}{Software/Hardware Engineering with the Parallel
  Object-Oriented Specification Language}}.
\newblock In: {\sl \bibinfo{booktitle}{Proceedings of MEMOCODE'07}},
  \bibinfo{publisher}{IEEE}, pp. \bibinfo{pages}{139--148},
  \doi{10.1109/MEMCOD.2007.371231}.

\bibitemdeclare{incollection}{T08}
\bibitem{T08}
\bibinfo{author}{J.~\surnamestart Tretmans\surnameend} (\bibinfo{year}{2008}):
  \emph{\bibinfo{title}{Model based testing with {Labelled Transition
  Systems}}}.
\newblock In: {\sl \bibinfo{booktitle}{Formal methods and testing}}, {\sl
  \bibinfo{series}{LNCS}} \bibinfo{volume}{4949},
  \bibinfo{publisher}{Springer}, pp. \bibinfo{pages}{1--38},
  \doi{10.1007/978-3-540-78917-8\_1}.

\bibitemdeclare{incollection}{VLHW13}
\bibitem{VLHW13}
\bibinfo{author}{J.~\surnamestart Verriet\surnameend}, \bibinfo{author}{H.L.
  \surnamestart Liang\surnameend}, \bibinfo{author}{R.~\surnamestart
  Hamberg\surnameend} \& \bibinfo{author}{B.~\surnamestart van
  Wijngaarden\surnameend} (\bibinfo{year}{2013}):
  \emph{\bibinfo{title}{Model-driven development of logistic systems using
  domain-specific tooling}}.
\newblock In: {\sl \bibinfo{booktitle}{Proceedings of CSD\&M}},
  \bibinfo{publisher}{Springer}, pp. \bibinfo{pages}{165--176},
  \doi{10.1007/978-3-642-34404-6\_11}.

\end{thebibliography}
\end{document}